# Instantaneous Photon Drag Currents in Topological Insulators

Yoshito Onishi[1, *], Zhi Ren[2], Mario Novak[2], Kouji Segawa[2], Yoichi Ando[2, †], & Koichiro Tanaka[1, 3, 4, ‡]

[1] *Department of Physics, Graduate School of Science, Kyoto University, Sakyo-ku, Kyoto 606-8502, Japan*

[2] *Institute of Scientific and Industrial Research, Osaka University, Ibaraki, Osaka 567-0047, Japan*

[3] *Institute for Integrated Cell-Material Sciences, Kyoto University, Sakyo-ku, Kyoto 606-8501, Japan*

[4] *CREST, Japan Science and Technology Agency, Kawaguchi, Saitama 332-0012, Japan*

**Topological insulator materials have been extensively studied in the field of condensed matter physics because nontrivial topology in the electronic state gives rise to a novel spin-polarized Dirac dispersion on the surface. To describe the electrodynamics of topological insulators, it is crucial to understand coherent and incoherent dynamics of carriers both in bulk and surface states. We applied terahertz emission spectroscopy to an intrinsic three-dimensional topological insulator material, $Bi_{1.5}Sb_{0.5}Te_{1.7}Se_{1.3}$, to elucidate ultrafast photo-induced carrier dynamics. The emitted terahertz electric field strongly depended on the polarization and incident angle of the excitation pulse. A three-fold rotational symmetry was clearly confirmed in the dependence of terahertz emissions on the azimuthal angle. The origin of terahertz emissions should be instantaneous photon drag currents induced by the excitation of femtosecond pulses.**

The nontrivial topology of the bulk insulating state results in a novel surface electronic state that is described by the massless Dirac equation with the spin state locked by the direction of momentum[1–11] in three-dimensional topological insulators (3DTIs). Electronic properties in the ground state have been investigated in the typical 3DTIs of $Bi_2Se_3$, $Bi_2Te_3$, $Sb_2Te_3$, and their solid solutions with angle-resolved photoemission spectroscopy (ARPES)[7–9], tunneling spectroscopy[10,11], and conductivity measurements[12–14]. To describe the electrodynamics of topological insulators, it is crucial to understand coherent and incoherent dynamics of carriers both in bulk and surface states. Their electric-field response such as the magneto-optical effect in the terahertz (THz) range[15] has recently been studied with optical pump-probe measurements[16–18] and time-resolved ARPES[19–21]. Time-resolved mid-infrared optical spectroscopy has recently revealed that Auger recombination is one of the main channels of population relaxation in the high carrier-density regime[22]. In contrast to a clear understanding of the carrier dynamics in picoseconds, ultrafast initial transients of the dynamics in sub-picoseconds have yet to be determined.

Measurement of photocurrent with a femtosecond laser pulse is a powerful method to know the nonequilibrium carrier dynamics in 3DTI materials. Recently, several groups have reported on the photocurrent between two electrodes attached to samples under the excitations of femtosecond laser pulses. McIver *et al.*[23] observed helicity-sensitive photocurrents in electron-doped 3DTIs, $Bi_2Se_3$, under near-infrared light illumination and phenomenologically assigned the origin to two kinds of photocurrents, i.e. photogalvanic current and photon drag current. Olbrich *et al.*[24] also observed photocurrent in 3DTIs, $Sb_2Te_3$, and $Bi_2Te_3$, under THz light illumination, but they assigned the main origin should be photogalvanic current, which is different from the previous report[23]. In addition, Duan *et al.*[25] have recently reported that helicity-sensitive photocurrent under near-infrared light illumination can be decomposed into two types of photocurrents originating from helical surface states and the surface accumulation layer. Controversial issues currently still remain on the origin while the mechanism for photocurrent in 3DTI materials has been attracting a great deal of interest.

Photocurrents in the experiments discussed above[23–25] were measured with electrodes fabricated on the sample surface, which resulted in difficulties with the experiments. The first issue was a lack of time resolution. It is difficult to conclude whether charge current is directly induced by light illumination without information on the transient response of the current, or indirectly by other effects such as the thermoelectric effect. The second issue is the fixed spatial relationship between the crystal axis and direction of current determined by two electrodes. Without the freedom of measuring the direction of current for given excitation conditions, it is difficult to claim the origin of photocurrents from the analysis of crystal symmetry based on group theory. One solution to overcome these difficulties is by applying THz emission spectroscopy. THz emission spectroscopy is a powerful tool to measure transient current in samples induced by femtosecond pulse excitation and to clarify the current dynamics in solids[26,27].

Here, we report THz emission spectroscopy on a 3DTI material, $Bi_{1.5}Sb_{0.5}Te_{1.7}Se_{1.3}$[13], under the illumination of femtosecond pulses with linear or circular polarizations. The laser pulses induced a transient charge current, $\boldsymbol{j}(t)$, through electronic excitation. The transient current may produce an electromagnetic wave in the far field, whose electric field is proportional to $\partial \boldsymbol{j}(t)/\partial t$, as is well known from the principles of classical electrodynamics. If current dynamics occurs in the sub-picosecond time region, one can expect an electromagnetic wave in the THz region that is nothing but a THz emission from the sample. The THz waveform enables us to extract dynamical information about the charge current. In addition, this contactless technique enables us to study the dependencies of the photocurrent on the crystal angle and the incident angle at high levels of reproducibility without having to attach electrodes. We observed a three-fold rotational symmetry for THz emissions produced by the photocurrent, which reflected crystal symmetry[23,28]. The dependence on the crystal angle and incident angle can well be attributed to the photon drag effect according to phenomenological tensor analysis. The time-domain analysis of the emitted THz waveforms indicates that photocurrent flowed instantaneously under light illumination. These results indicate that the illumination of femtosecond pulses induced instantaneous photon drag current in the 3DTI material.

Figure 1(a) has a schematic of the setup for THz emission spectroscopy. A single crystal of the 3DTI material $Bi_{1.5}Sb_{0.5}Te_{1.7}Se_{1.3}$ is irradiated by using a linearly or circularly polarized pulse that is the output from a Ti: sapphire oscillator (wavelength of 800 nm and pulse width of 100 fs). The horizontal and vertical components of the electric field were detected by using an electro-optical (EO) sampling technique with a 1-mm-thick ZnTe crystal. Figure 1(b) provides the crystal structure of $Bi_{1.5}Sb_{0.5}Te_{1.7}Se_{1.3}$ that is seen from the [111] direction ($z'$ axis). The crystal symmetry on the (111) surface is represented by the point group, $C_{3v}$, which consists of a three-fold rotational symmetry around the [111] axis and three mirror planes perpendicular to the (111) cleaved surface. We define a crystal coordinate system, $x'y'z'$, in which the $x'z'$ plane is one of the mirror planes and a laboratory coordinate system, $xyz$, in which the $z$ axis is identical to the $z'$ axis and the $xz$ plane is the incident plane [see Fig. 1(c)]. The azimuthal angle, $\theta$, is defined as the angle between the $x$ and $x'$ axes. The incident angle, $\psi$, is defined as the angle between the [111] axis and the wave vector $\boldsymbol{k}$ of incident light. THz emission spectroscopy by varying these two parameters, $\theta$ and $\psi$, provides enough information to assign the photocurrent. The direction of the $x'$ axis was determined on the basis of the azimuthal angle dependence of the emitted THz electric fields [see discussion on Fig. 3]. The detected THz electric field induced by right-handed circularly polarized light is denoted by $\boldsymbol{E}_R(t)$, that induced by left-handed circularly polarized light is denoted by $\boldsymbol{E}_L(t)$, that induced by vertically polarized light is denoted by $\boldsymbol{E}_V(t)$, and that induced by horizontally polarized light is denoted by $\boldsymbol{E}_H(t)$.

Figure 2(a) shows the THz emissions from $Bi_{1.5}Sb_{0.5}Te_{1.7}Se_{1.3}$ at an azimuthal angle of $\theta = 0°$ and an incident angle of $\psi = 22.5°$. Here, we decomposed the THz emissions induced by linearly polarized light into two components [see Supplementary Information 1]: linear-polarization-insensitive component $\langle E_l \rangle(t) = [E_V(t) + E_H(t)]/2$ and linear-polarization-sensitive component $\Delta E_l(t) = E_V(t) - E_H(t)$ [see Fig. 2(a)]. We similarly decomposed the THz radiations induced by the circularly polarized light into two components: helicity-insensitive component $\langle E_c \rangle(t) = [E_R(t) + E_L(t)]/2$ and helicity-sensitive component $\Delta E_c(t) = E_R(t) - E_L(t)$ [see Fig. 2(b)]. The components $\langle E_l \rangle(t)$ and $\langle E_c \rangle(t)$ have almost the same shape and magnitude, and are entirely horizontally polarized. Since these THz emissions are horizontally polarized and insensitive to the incident polarizations, we considered that $\langle E_l \rangle(t)$ and $\langle E_c \rangle(t)$ included contributions from photo-Dember or surface-depletion currents. Luo *et al.*[29] also reported the THz emissions produced by the bulk current in the direction normal to the surface and concluded that it originated from the photo-Dember or surface-depletion effects. Unlike Luo et al.[29], we observed a strong linear-polarization-sensitive ( $\Delta E_l(t)$ ) component in addition to polarization-insensitive THz emissions. In contrast to the observation of $\Delta E_l(t)$, we could observe no helicity-sensitive component $\Delta E_c(t)$ within the limits of our signal-to-noise ratio in any measurement setups with azimuthal angles $\theta$ and incident angles $\psi$. This is quite different from the observations by McIver et al.[23] and Duan et al.[25] We will discuss this point later.

To clarify the origin of the linear-polarization-sensitive ($\Delta E_l(t)$) components in more detail, we measured the dependence of emitted THz electric fields on the crystal angle to elucidate the mechanism responsible for photocurrent. Figure 3(a) plots the trajectories of linear-polarization-sensitive THz field $\Delta E_l(t)$ at $\theta = 0°$ (a1) and 30° (a2) detected with a lock-in technique synchronized with polarization modulation to improve the signal-to-noise ratio [see Supplementary Information 2]. This result clearly indicates that $\Delta E_l(t)$ does not simply originate from the difference in reflection loss. We modulated incident polarization by rotating the half-wave plate at a frequency of 10 Hz to improve the signal-to-noise ratio of the polarization-sensitive waveform and performed lock-in detection of the signal synchronized with this modulation frequency. As we can see from panel (a1), the emitted wave at $\theta = 0°$ is entirely horizontally polarized. In contrast, the THz emission at $\theta = 30°$ (a2) is entirely vertically polarized. These results clearly indicate that linearly directed photocurrent is driven by linearly polarized light, wherein the direction of the photocurrent rotates as the crystal rotates.

The detailed $\theta$ dependence of THz emissions provides important insights into this nontrivial behavior. Figure 3(b) plots the vertical (b1) and horizontal (b2) components of the linear-polarization-sensitive THz electric field at $t_0 = -0.13$ ps [indicated by the arrow in panel (a1) of Fig. 2(a)] plotted as a function of $\theta$ at an incident angle $\psi = 22.5°$. The effect of the oscillatory component after light irradiation can be neglected at negative delay. This arises from the group-velocity mismatch between the THz and sampling pulses in the electro-optic (EO) crystal. We can see that the linear-polarization-sensitive THz electric field follows the function

$$\Delta \boldsymbol{E}_l(t) = \boldsymbol{E}_\text{V}(t) - \boldsymbol{E}_\text{H}(t) = E_l^0(t)\begin{pmatrix}\cos 3\theta \\ \sin 3\theta\end{pmatrix}. \quad (1)$$

It apparently reflects the three-fold rotational symmetry of the crystal. These results indicate that the polarization of the emitted THz pulse rotates with $3\theta$ dependence. The amplitude, $E_l^0(t_0)$, of the variation corresponds to $3.2 \pm 0.2$ mV/cm.

Figure 4 plots the dependence of the horizontal component of $\Delta \boldsymbol{E}_l(t_0)$ on excitation light intensity $I$ at $\theta = 0°$ and $\psi = 22.5°$. It shows that the emitted THz electric field is proportional to light intensity [see the black line in Fig. 4]. The linear dependence on $I$ suggests that the emitted THz waves can be produced from second-order nonlinear photocurrents[30].

The dependence of THz emissions on the incident angle gives us more detailed information to determine the mechanism for the photocurrents. Here, we define conversion efficiency $\boldsymbol{\eta}_{V(H)}$ as the ratio of THz field $\boldsymbol{E}_{V(H)}(t_0)$ to absorbed light intensity, $I_{\text{ab},V(H)} = (1 - R_{V(H)})I$, where $1 - R_{V(H)}$ is the reflection loss for the vertically (horizontally) polarized light, estimated from the complex refractive index[22] to compensate for the difference in reflection loss between incident polarizations. Figure 5 shows $\Delta \boldsymbol{\eta}_l = \boldsymbol{\eta}_V - \boldsymbol{\eta}_H$ [see panel (a)] and $\langle \boldsymbol{\eta}_l \rangle = [\boldsymbol{\eta}_V + \boldsymbol{\eta}_H]/2$ [see panel (b)], plotted as a function of $\psi$. We found that the horizontal component of $\Delta \boldsymbol{\eta}_l$ at $\theta = 180°$ [see panel (a1)] and the vertical component of $\Delta \boldsymbol{\eta}_l$ at $\theta = 30°$ [see panel (a2)] were increasing with the angle of incidence and were largest around $\psi = 60°$. Also, the horizontal component of $\langle \boldsymbol{\eta}_l \rangle$ at $\theta = 180°$ [see panel (b1)] was dependent on incident angle $\psi$ and increased with $\psi$, which suggests that the linear-polarization-insensitive component of the emitted THz field $\langle \boldsymbol{E}_l \rangle(t)$ should not only be derived from the photo-Dember or surface depletion effect[29], but also from another effect. We also measured the vertical components of $\langle \boldsymbol{\eta}_l \rangle$, which were always small even at the large $\psi$ [see panel (b2)].

Second-order nonlinear photocurrents can be written as

$$j^\lambda(t,\boldsymbol{r}) = \sum_{\nu,\eta=x,y,z} \sigma^\lambda{}_{\nu\eta} \mathcal{E}(t,\boldsymbol{r})^\nu (\mathcal{E}(t,\boldsymbol{r})^*)^\eta + i \sum_{\mu,\nu,\eta=x,y,z} \Phi^\lambda{}_{\mu\nu\eta} \mathcal{E}(t,\boldsymbol{r})^\nu \frac{\partial(\mathcal{E}(t,\boldsymbol{r})^*)^\eta}{\partial x_\mu} \quad (2)$$

$$+ c.c.,$$

where the first term is categorized as the photogalvanic effect and the second term is categorized as the photon drag effect[30]. Here, $\sigma^\lambda{}_{\nu\eta}$ is called the photogalvanic tensor, $\Phi^\lambda{}_{\mu\nu\eta}$ is called the photon drag tensor, and $c.c.$ means the complex conjugate of all preceding terms in the expression. $\mathcal{E}(t,\boldsymbol{r})^\nu$ is the electric field of incident light. For example, the electric fields of vertically and horizontally polarized lights correspond to $\mathcal{E}(t,\boldsymbol{r})^\nu = \mathcal{E}_0(0,1,0)e^{i\boldsymbol{k}\cdot\boldsymbol{r}-i\omega t}$ and $\mathcal{E}(t,\boldsymbol{r})^\nu = \mathcal{E}_0(-\cos\psi, 0, \sin\psi)e^{i\boldsymbol{k}\cdot\boldsymbol{r}-i\omega t}$, where $\mathcal{E}_0$ is the amplitude of the incident electric field, and $\omega$ is the frequency of incident light. In the following, we discuss the photocurrent mechanism through phenomenological analysis.

The nonvanishing elements of the photogalvanic tensor, $\sigma^\lambda{}_{\nu\eta}$, and the photon drag tensor, $\Phi^\lambda{}_{\mu\nu\eta}$, can be determined from crystal symmetry invariance for transformation under $C_{3v}$ rotational and mirror symmetry operations [see Supplementary Information 3 and Refs.[18,30]]. It is clearly shown that the linear-polarization-sensitive component of the photogalvanic current has a dependence of $1 + \cos^2\psi$, which becomes largest at $\psi = 0°$ and decreases monotonously with incident angle $\psi$. This behavior is inconsistent with our observed incident-angle dependence as shown in Fig. 5. Hence, we can rule out the photogalvanic effect as the origin of photocurrent.

In contrast, photon drag currents can be described by

$$\Delta \boldsymbol{j}_l(t) = kI\left[-\left(d^{(1)} + d^{(2)}\sin^2\psi\right)\cos\psi\begin{pmatrix}\cos 3\theta \\ \sin 3\theta\end{pmatrix} + \left(d^{(3)} - d^{(4)}\sin^2\psi\right)\sin\psi\begin{pmatrix}1 \\ 0\end{pmatrix}\right], \qquad (3)$$

for the linear-polarization-sensitive component, and

$$\langle \boldsymbol{j}_l\rangle(t) = \frac{kI}{2}\left[d^{(2)}\sin^2\psi\cos\psi\begin{pmatrix}\cos 3\theta \\ \sin 3\theta\end{pmatrix} + \left(-d^{(5)} + d^{(4)}\sin^2\psi\right)\sin\psi\begin{pmatrix}1 \\ 0\end{pmatrix}\right], \qquad (4)$$

for the linear-polarization-insensitive component, where $k = |\boldsymbol{k}|$ and $d^{(i)}$ ($i = 1,\cdots,5$) are the linear combination of nonvanishing elements of the photon drag tensor. The first term in Eq. (3) indicates three-fold rotational symmetry and its dependence on the incident angle indicates the local minimum value at $\psi = 0°$, depending on the values of the photon drag tensor, $d^{(1)}$ and $d^{(2)}$. This dependence of photon drag current on the incident angle can be in good qualitative agreement with the experimental dependence on the incident angle as will be explained in what follows. Note that this $3\theta$ dependence is seemingly quite a different form from the previously reported $2\theta$ dependence[24]. This difference simply stems from the coordinate system in the experimental setup.

Here we have used the laboratory coordinate system $xyz$, which reflects the crystal three-fold rotational symmetry in more obvious manner.

Least squares fitting analysis has been made with eqs. (3) and (4) on the incident angle. Here we added a constant term $d^{(0)}$ in the fitting for the horizontal component of $\langle \boldsymbol{\eta}_l \rangle$, which corresponds to the photo-Dember or surface depletion current that only appears in the linear-polarization-insensitive component. The best-fitted curves are shown by the solid lines in Fig. 5 and in good agreement with all components of the experimental $\psi$-dependence. We determined the relative values of all elements of the photon drag tensor, $d^{(i)}$ ($i = 1, \cdots, 5$), from the fitting results normalized with the photo-Dember or surface depletion contribution $d^{(0)}$ as $d^{(1)}/d^{(0)} = 0.31 \pm 0.03$, $d^{(2)}/d^{(0)} = 1.17 \pm 0.09$, $d^{(3)}/d^{(0)} = 0.09 \pm 0.11$, $d^{(4)}/d^{(0)} = 0.1 \pm 0.2$, and $d^{(5)}/d^{(0)} = 3.4 \pm 0.3$.

The time-domain analysis of the THz waveforms enabled us to understand the transient response of photocurrent. Deconvolution analysis [see Supplementary Information 4] indicates clearly that linear-polarization-sensitive photocurrent has an almost instantaneous response and its relaxation time should be faster than the excitation pulse width of 100 fs, which suggests a strong momentum scattering process in the very beginning of relaxation. The fast relaxation time of the photocurrent can originate from many-body scattering under the high-density excitation, such as the Auger recombination[22]. We may be able to identify why we could not observe any helicity-sensitive THz emissions, as shown in Fig. 2(b) due to transient dynamics. The lack of helicity-sensitive photocurrent is quite different from the previous observations of photocurrent with time-integrated techniques[23,25]. This suggests that helicity-sensitive photocurrent flows in a longer timescale after the initial relaxation has been completed. Such photocurrent with long lifetimes could not be detected with THz emission spectroscopy that covers subpicosecond to picosecond time scales, but was detected with time-integrated measurements. This suggests incoherent origin of the helicity-sensitive photocurrent not driven by the light field directly. Another possible reason of the lack of helicity-sensitive photocurrent is the absence of the surface carriers before photoexcitation. To reveal it, we should study the detailed dependence on the Fermi levels of the emitted THz fields.

It is noteworthy to discuss whether photon drag current stemmed from bulk or from surface current. Since the photon drag effect is allowed even in centrosymmetric media, photon drag current may not only flow on the surface but also in bulk. In fact, we cannot rule out the possibility that current flows in bulk only from our results. Olbrich *et al.*[24] concluded that photocurrent originated from the semi-classical motion of Dirac surface electrons under low-frequency excitation in the THz region. Since the current derives from intraband excitation in Dirac dispersion, this contribution should decrease drastically with increasing excitation frequency as was explained by Olbrich *et al.*

[24]. This suggests that this contribution is negligible for near-infrared excitation. To clarify this, we should address the microscopic origin of the photon drag effect more in the future.

In conclusion, we first observed the linear-polarization-sensitive component of the THz emissions from an intrinsic 3DTI material, $Bi_{1.5}Sb_{0.5}Te_{1.7}Se_{1.3}$, under the illumination of a linearly polarized femtosecond pulse. Details on the dependence of THz emissions on the azimuthal angle, excitation intensity, and incident angle revealed that photocurrent that induced THz emissions should originate from a second-order photon drag effect. Time-domain analysis discovered that photocurrent flowed instantaneously and its relaxation time was surprisingly fast. Although understanding of the microscopic origins requires more careful theoretical treatment, the discovery of novel THz emissions in 3DTIs presents a new angle in addressing interesting nonequilibrium physics.

**Methods**

**Sample Preparation:** High-quality single crystals of $Bi_{1.5}Sb_{0.5}Te_{1.7}Se_{1.3}$ were grown by melting stoichiometric amounts of high-purity elements in sealed quartz tubes at 850℃ for 48 h and then cooling them slowly to 550℃, at which the boule was kept for 4 d to reduce crystals defects. X-ray diffraction analyses confirmed that the crystals had a chalcogen-ordered tetradymite structure. Representative crystals cut from the boules of $Bi_{1.5}Sb_{0.5}Te_{1.7}Se_{1.3}$ used for this study were characterized by measuring the dependence of resistivity and the low-temperature Hall coefficient on temperature, which were confirmed to be similar to those that have been reported[9,12–14].

**Experimental Setup:** The light source was a Ti: sapphire oscillator producing near-infrared femtosecond light pulses (wavelength of 800 nm, repetition rate of 82 MHz, and pulse width of 100 fs). The linearly polarized light pulse was incident to a rotating half-wave plate and a quarter-wave plate and was converted into each polarization state including linear or circular polarization depending on the angle of rotation of the half-wave plate and quarter-wave plate. The converted light was focused on the cleaved surface of a single crystal of the 3DTI material, $Bi_{1.5}Sb_{0.5}Te_{1.7}Se_{1.3}$, with a spot size of 30 μm in diameter. The generated emissions were focused onto a 1-mm-thick ZnTe crystal, after the near-infrared excitation pulse was removed by using a silicon plate. The horizontal and vertical components of the electric field were detected by using an electro-optical (EO) sampling technique with two wire-grid polarizers, the extinction ratio of which was ~$10^{-4}$ below 7 THz. As the downstream polarizer was fixed, only polarization tilted by 45° from the horizontal axis could pass through. Either component of the emitted wave could be detected without rotating the ZnTe crystal by either rotating the upstream polarizer horizontally or vertically. All measurements were carried out in dry air at room temperature.

**Acknowledgements**

Y.O. thanks T. Arikawa for his help in constructing the optical setup and S. Tani for stimulating discussions. K.T. acknowledges support by a Grant-in-Aid for Scientific Research on Innovative Areas, titled 'Optical science of dynamically correlated electrons (DYCE)' (Grant No. 20104007), a Grant-in-Aid for Scientific Research (A) (Grant No. 26247052), and a Grant-in-Aid for the Global COE Program 'The Next Generation of Physics, Spun from Universality and Emergence 'from the MEXT of Japan. Y.A. acknowledges support by a Grant-in-Aid for Scientific Research on Innovative Areas, titled 'Topological Quantum Phenomena' (Grant No. 22103004), a Grant-in-Aid for Scientific Research (S) (Grant No. 25220708), and the *Air Force Office of Scientific Research of the USA* (Grant No. AOARD 124038).

**Author Contributions**

Y.O. developed the optical system, performed the measurements, analyzed the data, developed the theoretical framework, and wrote the paper together with K.T. and Y.A. K.T. supervised the measurements in the project and Y.A. supervised the materials. Z.R., M.N., and K.S. carried out the growth and characterizations of the samples.

**Competing Interests**

The authors declare that they have no competing financial interests.

**Correspondence**

Correspondence and requests for materials should be addressed to Y.O. (email: onishi@scphys.kyoto-u.ac.jp), Y.A. (email: y_ando@sanken.osaka-u.ac.jp), or K.T. (email: kochan@scphys.kyoto-u.ac.jp).

016401 (2010).

11. Cheng, P. *et al.* Landau Quantization of Topological Surface States in $Bi_2Se_3$. *Phys. Rev. Lett.* **105,** 076801 (2010).

12. Segawa, K. *et al.* Ambipolar transport in bulk crystals of a topological insulator by gating with ionic liquid. *Phys. Rev. B* **86,** 075306 (2012).

13. Ren, Z., Taskin, A. A., Sasaki, S., Segawa, K. & Ando, Y. Optimizing $Bi_{2-x}Sb_xTe_{3-y}Se_y$ solid solutions to approach the intrinsic topological insulator regime. *Phys. Rev. B* **84,** 165311 (2011).

14. Taskin, A. A., Ren, Z., Sasaki, S., Segawa, K. & Ando, Y. Observation of Dirac Holes and Electrons in a Topological Insulator. *Phys. Rev. Lett.* **107,** 016801 (2011).

15. Valdés Aguilar, R. *et al.* Terahertz Response and Colossal Kerr Rotation from the Surface States of the Topological Insulator $Bi_2Se_3$. *Phys. Rev. Lett.* **108,** 087403 (2012).

16. Hsieh, D. *et al.* Nonlinear Optical Probe of Tunable Surface Electrons on a Topological Insulator. *Phys. Rev. Lett.* **106,** 057401 (2011).

17. Hsieh, D. *et al.* Selective Probing of Photoinduced Charge and Spin Dynamics in the Bulk and Surface of a Topological Insulator. *Phys. Rev. Lett.* **107,** 077401 (2011).

18. McIver, J. W. *et al.* Theoretical and experimental study of second harmonic generation from the surface of the topological insulator $Bi_2Se_3$. *Phys. Rev. B* **86,** 035327 (2012).

19. Sobota, J. A. *et al.* Ultrafast Optical Excitation of a Persistent Surface-State Population in the
12

**Figure Legends**

**Figure 1| Generation of terahertz waves from topological insulators excited by linearly and circularly polarized light.** (a) Schematic of experimental setup to detect terahertz waves excited by pulses of linearly and circularly polarized light. (b) Crystal structure of 3DTI material, $Bi_{1.5}Sb_{0.5}Te_{1.7}Se_{1.3}$, as seen from [111] direction. (c) Relation between crystal- and laboratory-coordinate systems.

**Figure 2| Generated terahertz waveforms from $Bi_{1.5}Sb_{0.5}Te_{1.7}Se_{1.3}$.** (a) Electric fields of average $\langle E_l \rangle(t) = (E_V(t) + E_H(t))/2$ and difference $\Delta E_l(t) = E_V(t) - E_H(t)$ and (b) average $\langle E_c \rangle(t) = (E_R(t) + E_L(t))/2$ and difference $\Delta E_c(t) = E_R(t) - E_L(t)$ as a function of delay time, emitted from $Bi_{1.5}Sb_{0.5}Te_{1.7}Se_{1.3}$ at azimuthal angle $\theta = 0°$, incident angle $\psi = 22.5°$, and intensity $I = 0.34$ mJ/cm$^2$. All measurements were done at room temperature. For convenience, time origin is defined as zero-crossing point of detected terahertz waveform. Panels (a1) and (b1) indicate horizontal components and panels (a2) and (b2) indicate vertical components. Arrow in panel (a1) indicates time $t = -0.13$ ps, at which electric fields are plotted in Fig. 3b.

**Figure 3 | Dependence of electric field on azimuthal angle $\theta$ observed for $Bi_{1.5}Sb_{0.5}Te_{1.7}Se_{1.3}$.** (a) Trajectories of difference $\Delta E_l(t) = E_V(t) - E_H(t)$ at azimuthal angles $\theta = 0°$ (a1) and $30°$ (a2) and at incident angle $\psi = 22.5°$. Dashed blue lines indicate mirror planes on $Bi_{1.5}Sb_{0.5}Te_{1.7}Se_{1.3}$. (b) $\theta$-dependencies of vertical (b1) and horizontal (b2) components of linear-polarization-sensitive terahertz electric fields at $t = -0.13$ ps (circles) and their least-square fitting curves with sine functions (curves).

**Figure 4 | Dependence of electric field on light intensity observed for $Bi_{1.5}Sb_{0.5}Te_{1.7}Se_{1.3}$.** Dependence of observed electric-field amplitude $E_l^0(t_0)$ on light intensity $I$ (circles) and their least-square line fitting (black line).

**Figure 5 | Dependence of electric field on incident angle observed for $Bi_{1.5}Sb_{0.5}Te_{1.7}Se_{1.3}$.** Incident-angle dependencies of conversion efficiencies, $\Delta\eta_l$(a) and $\langle\eta_l\rangle$ (b), at azimuthal angle $\theta = 180°$(a1, b1) and $30°$(a2, b2). Panels (a1) and (b1) indicate horizontal components and panels (a2) and (b2) indicate vertical components. Circles are experimentally obtained conversion efficiencies, and black lines are best-fitting functions with incident-angle dependence of photon drag current in Eqs. (3) and (4).

**Fig. 1**

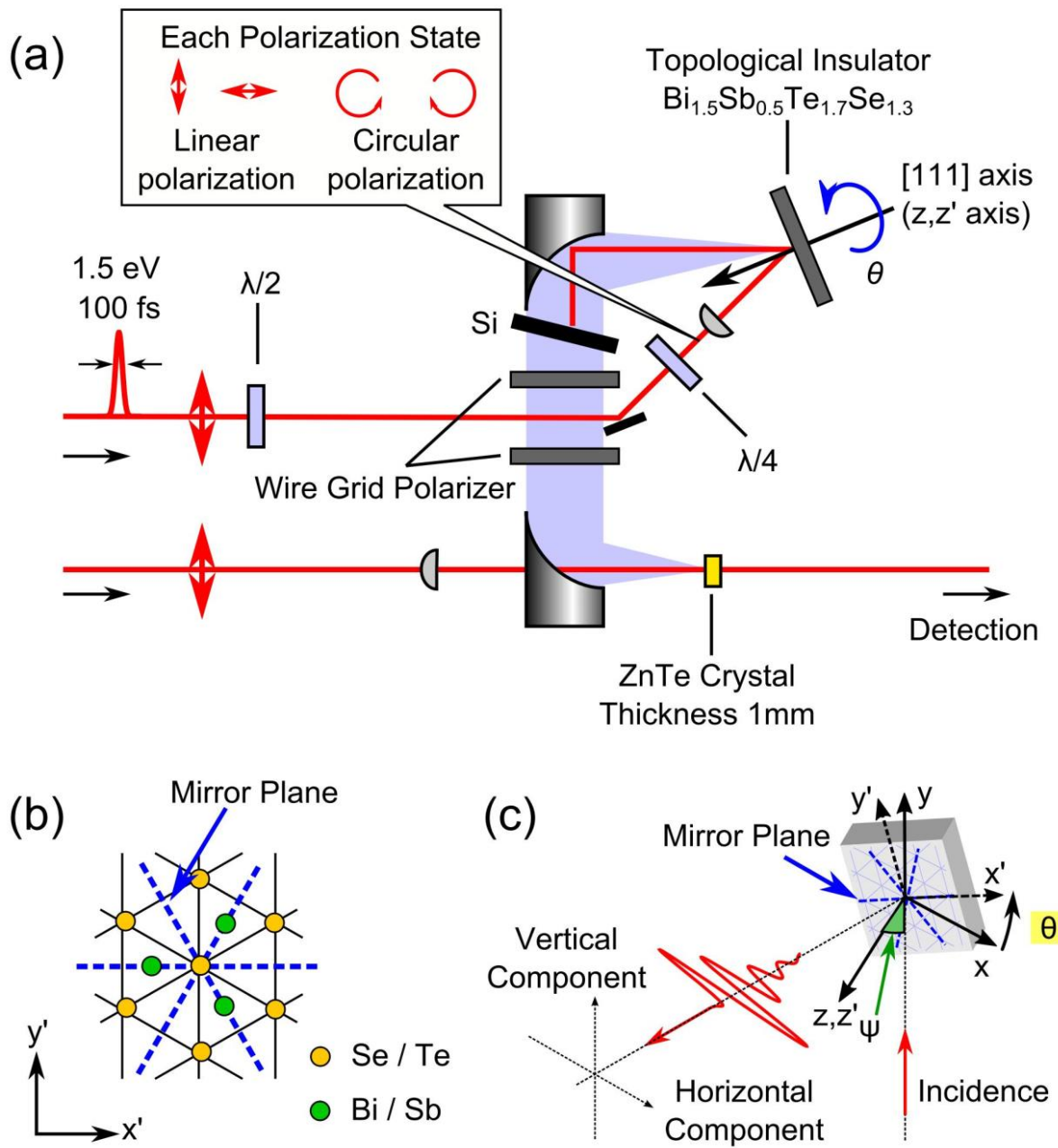

Y. Onishi et al.

**Fig. 2**

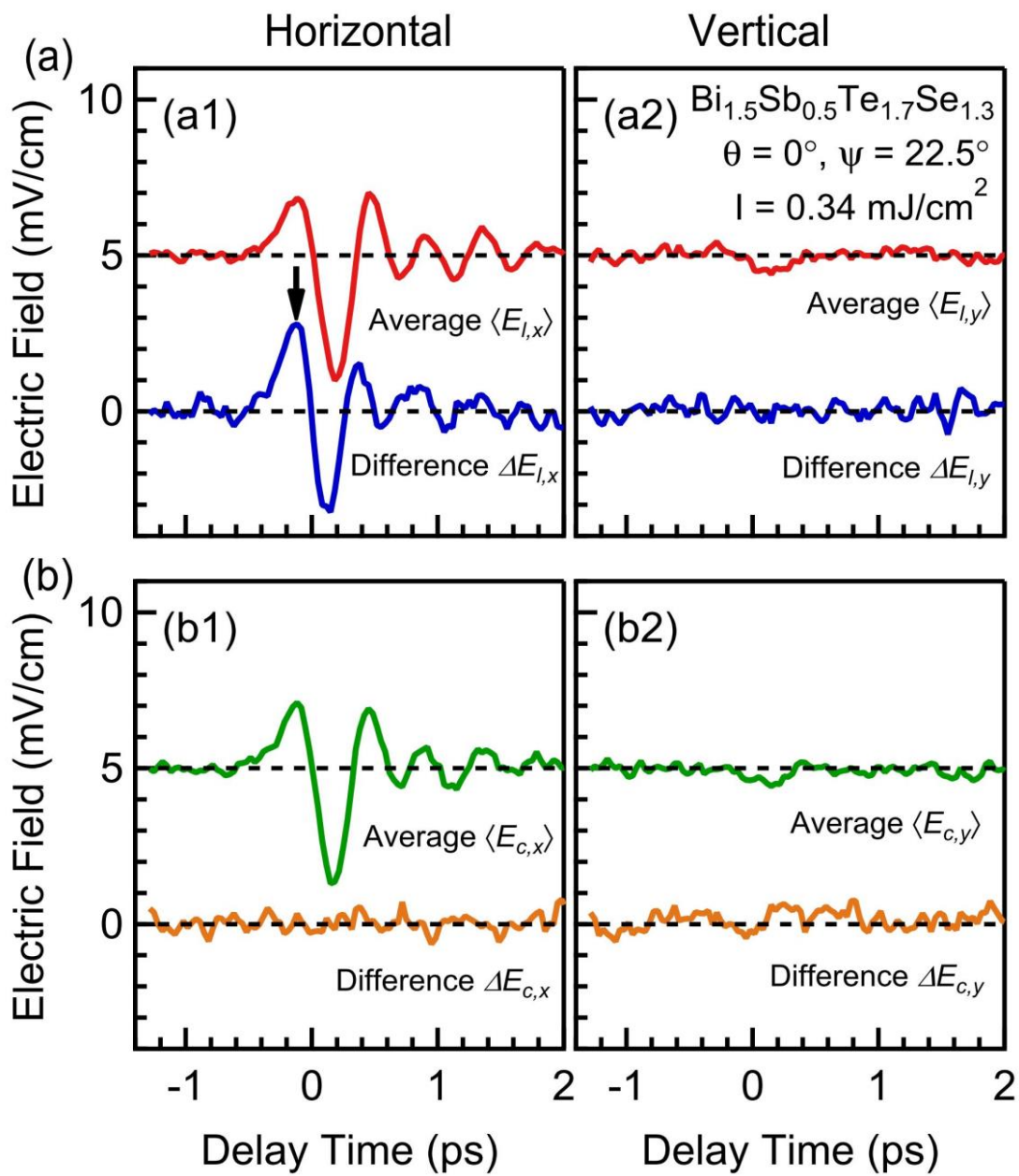

Y. Onishi et al.

**Fig. 3**

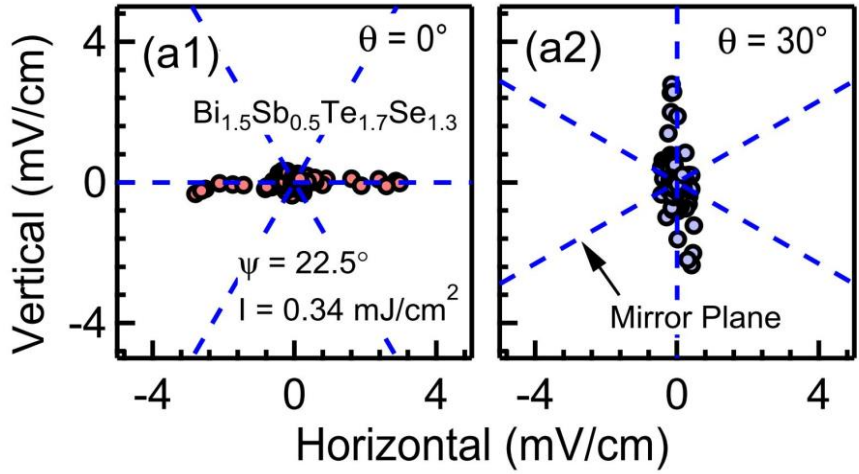

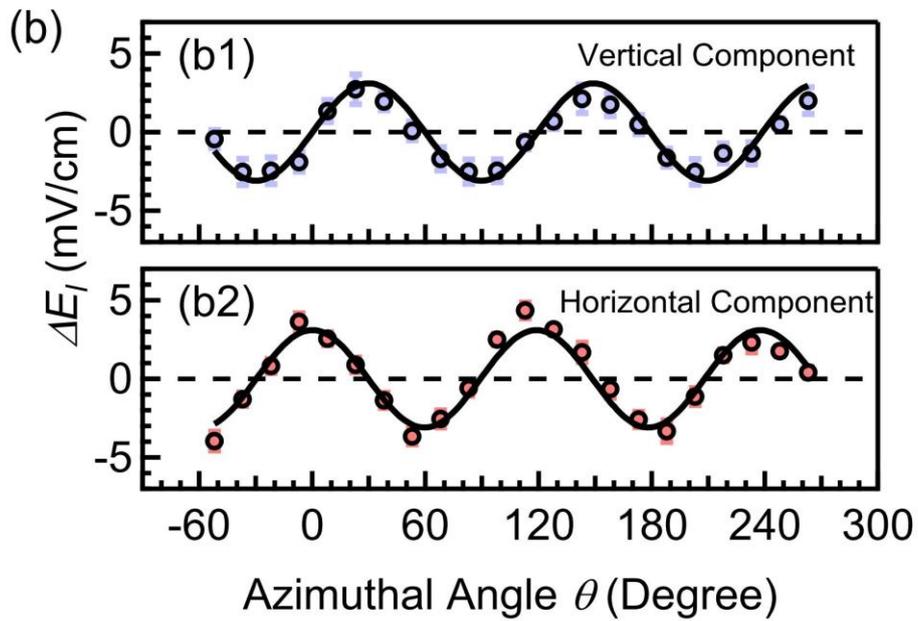

Y. Onishi et al.

**Fig. 4**

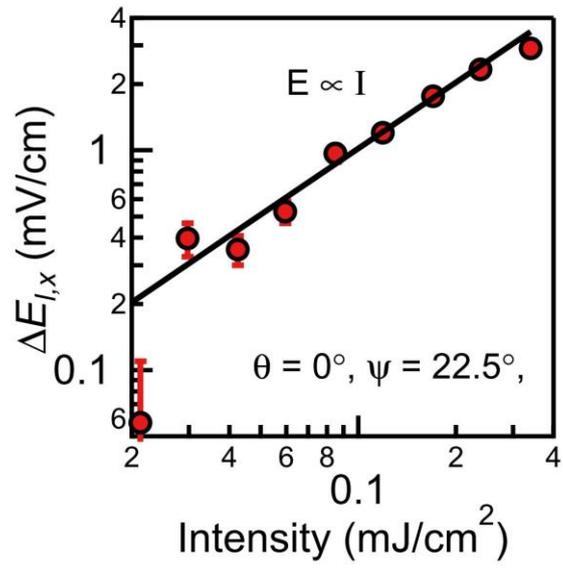

**Y. Onishi et al.**

**Fig. 5**

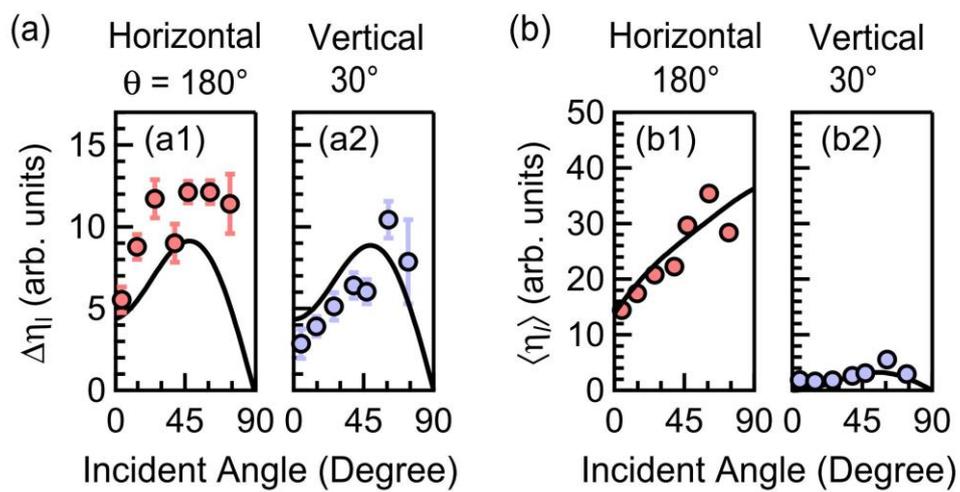

Y. Onishi et al.